\documentstyle[preprint,aps]{revtex}
\tightenlines
\begin{document}
\draft
\title{Structure and far-infrared edge modes of quantum
antidots at zero magnetic field}
\author{A. Emperador$^{1}$, M. Pi$^1$, M. Barranco$^1$,
E. Lipparini$^{2}$, and Ll. Serra$^3$.}
\address{$^1$Departament d'Estructura i Constituents de la Mat\`eria,
Facultat de F\'{\i}sica, \\
Universitat de Barcelona, E-08028 Barcelona, Spain}
\address{$^2$Dipartimento di Fisica, Universit\`a di Trento. 38050
Povo, Italy}
\address{$^3$Departament de F\'{\i}sica, Facultat de Ci\`encies,\\
Universitat de les Illes Balears, E-07071 Palma de Mallorca, Spain}
\date{\today}

\maketitle

\begin{abstract}

We have investigated  edge modes of different
multipolarity sustained by quantum antidots at zero
magnetic field.  The ground state of the antidot is described within
a local density functional formalism. Two sum rules, which are exact
within this formalism,  have been derived and used to evaluate the
energy of edge collective modes as a function of the surface density
and the size of the antidot.

\end{abstract}


\narrowtext
\section*{}

\section*{}

With the progress of microstructure technology, the study of the two
dimensional electron gas (2DEG)
has evolved to that of laterally confined
superlattices made of either electron islands (dots) or holes
surrounded by electrons (antidots). Much effort has been devoted in
the past to the study of quantum dots, as compared to that put in
the study of quantum antidots. One of the goals of
their study has been to determine the far-infrared response of these
semiconductor microstructures, and the formation of
compressible and incompressible states when a magnetic field $B$ is
perpendicurlaly applied. In the case of antidots,
which is the subject of the present work, experimental evidence of
collective excitations sustained by these structures has been
presented in Refs. \cite{ke91,zh92,lo92}. A
theoretical description based on magnetoplasmons in two-dimensional
antidot structures has been given in Ref. \cite{wu93} which compares
well with the experimental data of Ref. \cite{zh92}. Recently, the
existence of compressible and incompressible strips at the edge of
antidots has been determined by far-infrared spectroscopy \cite{bo96}.

We have started a systematic study of the
structure and collective far-infrared response of antidots, whose
aim is to achieve a level of sophistication
in the description of these systems
similar to that attained for quantum dots. As a first
step, we present here results at zero magnetic field obtained within
Density Functional Theory (DFT). To some extend, the present study is
similar in scope to that carried out in Ref. \cite{se92} on the surface
excitations of cavities in 3D metals. The $B \neq$ 0 case, which
requires a rather different and far more complex approach, will be
presented in a forthcoming paper.

We have modelled an antidot of radius $R$ in a 2DEG of surface
density $n_s$ by a positive jellium background of density
$n_s\,\Theta (r - R)$ to which we have added for the reason
that will be given below, a parabolic potential barrier
of the type $V_p(r)= m^* \omega_0^2 (R^2-r^2)/2$ acting on the
electrons for $r \leq R$.
We shall call $V_{ext}$ the sum of the jellium
and $V_p$ potentials.  The ground state (gs) of the antidot is
obtained solving the Kohn-Sham (KS) equations as indicated for example
in Ref. \cite{fe94}. The problem is simplified by the
imposed circular symmetry, and only the radial KS equations have to
be considered to determine the  electronic radial wave functions
$R_l(r)$:
\begin{equation}
\frac{d^2 R_l(r)}{d r^2} + \frac{1}{r} \frac{d R_l(r)}{d r} +
\left[ \frac{2 m^*}{\hbar^2} \left( E - V(r) \right) -
\frac{l^2}{r^2} \right] R_l(r) = 0 \,\,\, ,
\label{eq1}
\end{equation}
where m$^*$ is the electron effective mass which together with a
dielectric constant $\epsilon$ are
characteristics of the semiconductor. For example,
$\epsilon$ = 12.4 and m$^*$ = 0.067 m$_e$ in GaAs, which we have chosen
for the numerical applications in view of the existing experimental
data of Refs. \cite{zh92,bo96}. $l = 0, \pm 1, \pm 2$...
is the angular momentum about the $z$ axis perpendicular to the plane
of the antidot, and $r$ is the radial variable in that plane. The
single electron potential $V(r)$ is made of $V_{ext}$, of the
Hartree electron-electron potential and of the exchange-correlation
potential. The correlation potential has been obtained from the
correlation energy of Ref. \cite{ta89} in a local density
approximation.
Some times we shall use effective atomic units. In this system
of units, the length unit is the Bohr radius times $\epsilon/m^*$,
and the energy unit is the Hartree times $m^*/\epsilon^2$, denoted here
as $a_0^*$ and $H^*$, respectively.
$\epsilon$ is the dielectric constant, and
$m=m^* m_e$ is the electron effective mass.
For GaAs we have
$\epsilon$=12.4 and $m^*$=0.067. Consequently, $a_0^* \sim
97.9 {\rm \AA}$ and $H^* \sim $ 11.9 meV $\sim$ 95.6 cm$^{-1}$.

Physically acceptable solutions to Eq. (\ref{eq1}) have to be
regular at $r = 0$ and behave asymptotically as
$R_l(r,k) \sim J_l(kr) + tan (\delta_l) Y_l(kr)$,
where $J_l$ and $Y_l$ are the integer Bessel functions of first and
second kind \cite{ab70}, and $k = \sqrt{2 m^* (E - V_{\infty})
/ \hbar^2}$. The wave number $k$ has to be
$k \le k_F = \sqrt{2 \pi n_s}$. Taking into consideration the
spin degeneracy, the electron density $\rho(r)$ is obtained as
\begin{equation}
\rho(r) = \sum_{l=-\infty}^{l=\infty} \frac{2}{(2 \pi)^2}
\int_{|\vec{k}| \leq k_F}  R_l^2(r,k) \,d \,\vec{k}  \,\,\, .
\label{eq2}
\end{equation}
We have checked that the number of points used to carry out a Gaussian
integration over $k$, and the maximum $|l|$ employed  in Eq.
(\ref{eq2}) lead to stable results. Typically, some 1000-1500 wave
functions have been computed to obtain a density.

We show in Fig. (\ref{fig1}) the electronic densities
corresponding to antidots of
$R$ = 7.5 $a_0^*$ and $n_s$ = 0.05, 0.1, 0.2, 0.3 and 0.4
$(a_0^*)^{-2}$.
They span the size and density range of
those fabricated and experimentally studied in
Ref. \cite{zh92}, and have been obtained using
$\omega_0 = 0.09 H^*$.
Figure (\ref{fig6}) shows the $V(r)$ KS potential
corresponding to $n_s$ = 0.1 and 0.4 $(a_0^*)^{-2}$. Also indicated
are the Fermi energy and the contribution of $V_p(r)$ to $V(r)$.

Looking at the position of the Fermi energy with respect of the top of
$V(r)$, it is not surprising that a potential barrier of
a kind or another is needed to prevent the electrons from spilling in
the antidot too much, producing an unphysical representation of the
experimental device. Actually, we have found that if $\omega_0$
is set to zero, high density antidots would have
non-zero electron densities even at $r$ = 0. This is illustrated in
Fig. (\ref{fig2}) for the $R$ = 180 nm, $n_s = 9 \times 10^{11}$
cm$^{-2}$ antidot of Ref. \cite{bo96}. In this figure, the dashed line
density has been obtained setting $\omega_0$ = 0.

These results can be employed to determine the $B$ = 0
far-infrared multipole response of antidots. To do so, we
rely on the formalism described in detail in Ref. \cite{li97}. For the
present purpose, it consists in obtaining the so-called
$m_1$ and $m_3$ sum rules (SR) for an excitation operator $Q_L$.
We have that \cite{li89}:
\begin{eqnarray}
m_1(Q_L) & = & \frac{1}{2} \langle 0| [Q_L^+,[H,Q_L]] |0 \rangle
\nonumber
\\
& &
\label{eq3}
\\
m_3(Q_L) & = & \frac{1}{2} \langle 0| [[H,[H,Q_L^+]],[H,Q_L]]
|0 \rangle
\,\, ,
\nonumber
\end{eqnarray}
where $|0 \rangle$ is the gs of the system.
These SR have been extensively studied in the literature
\cite{li89,bo79}. As indicated in these references, if only a
collective state is contributing to the strenght function,
 $E_3(Q_L) \equiv (m_3/m_1)^{1/2}$ represents the
average excitation energy. This is the situation experimentally
found for antidots at zero magnetic field.

The operator $Q_L$ is taken to be
\begin{equation}
Q_L=\sum^N_{j=1} \frac{1}{r^L_j} \, e^{iL\theta_j}
\,\,\,.
\label{eq4}
\end{equation}
This choice is inspired in  that $(qr)^{-L} e^{iL\theta}$ is the small
$q$ expansion of the function $Y_L(qr) e^{iL\theta}$, which is the
restriction to
the $z$ = 0 plane of the irregular solution of the Laplace equation in
cylindrical coordinates.
Following Ref. \cite{li97}, a lengthy
but straightforward calculation yields:
\begin{equation}
m_1(Q_L)= 2 \pi L^2 \int_0^{\infty} \,d r\, \frac{1}{r^{2L+1}}
\,\rho(r)
\label{eq5}
\end{equation}
\begin{equation}
m_3(Q_L)=m_3(T) + m_3(ee) + m_3(V_{ext}\,e) \,\,\, ,
\label{eq6}
\end{equation}
where
\begin{equation}
m_3(T)  =  2 \pi L^2(L+1) \int_0^{\infty} d r \,
\frac{1}{r^{2L+3}}
 \left[L \tau(r) + 2(L+2) \lambda(r) \right]
\label{eq7}
\end{equation}
\begin{eqnarray}
m_3(ee) & = & 4 \pi L^2\frac{(2L-1)!!}{2^L\,L!}
\int_0^\infty  \rho'(r)dr \left\{\frac{1}{r^{2L+1}}
\int^r_0 \left[ 2 \rho'(r') +r' \rho''(r') \right]
E_L\left(\frac{r'}{r}\right) dr' \right.
\nonumber
\\
& + & \left. \int^\infty_r \frac{1}{r'^{(2L+1)}}
\left[(2L+1)\rho'(r')
-r'\rho''(r')\right]E_L\left(\frac{r}{r'}\right)
dr'- \frac{2^{L+1} L!}{(2L+1)!!} \frac{1}{r^{2L}} \rho'(r)\right\}
\nonumber
\\
& + & 4 \pi L^2\int^\infty_0 dr\, \frac{1}{r^{2L+2}}
\left[r \rho''(r) - (2L+1) \rho'(r)\right]
\left\{\frac{1}{r}\int^r_0
\left[3  r'\rho(r')+ r'^2 \rho'(r')\right]
{\bf E}\left(\frac{r'}{r}\right)\,dr'\right.
\nonumber
\\
& - & \left. \int^\infty_r
r' \rho'(r') {\bf E} \left(\frac{r}{r'}\right) \,dr'
- 2 \,r \rho(r)
+ \lim_{R_{\infty} \rightarrow \infty} R_{\infty} n_s  {\bf E}
\left(\frac{r}{R_\infty}\right)  \right\}
\label{eq8}
\end{eqnarray}
\begin{equation}
m_3(V_{ext}\,e)=\pi L^2
\int^\infty_0 dr V_{ext}(r)\left[
\rho''(r)-\frac{2L+1}{r} \rho'\right] \frac{1}{r^{2L+1}} \,\, .
\label{eq9}
\end{equation}
The definition of the densities $\tau(r)$ and $\lambda(r)$ and
of the function $E_L$ can be found in Ref. \cite{li97}, and the
primes on the density denote $r$-derivatives. The jellium
 potential $V_+(r)$ entering in $V_{ext}$ is:
\begin{equation}
V_+(r)  =  4 n_s  \left\{ \begin{array}{ll}
R_{\infty}\,{\bf E}(r/R_{\infty})
-R \,{\bf E}(r/R) &  r < R  \\
R_{\infty}\,{\bf E}(r/R_{\infty}) - r\, {\bf E}(R/r)
+ r \,( 1 - (R/r)^2\,) \,{\bf K}(R/r)
 &     r > R  \,\, .
                \end{array}
\right.
\label{eq10}
\end{equation}
In the above equations, {\bf K} and {\bf E} are the complete
elliptic integrals of first and second kind, respectively
\cite{gr80}, and $R_{\infty}$ represents a large $r$ value.
In practice, it
is the largest $r$ used in the structure calculation, which
we have also taken as the point where the asymptotic behavior of
$R_l(r)$ holds. We want to point out that the two Coulomb
diverging terms in $m_3(ee)$ and $m_3(V_{ext}\,e)$ cancel each other.

The present formalism can be applied to antidots with the restriction
that $\rho(r)$ vanishes in a small disk around the origin.
In practice, this is not a limitation,
as can be inferred from Figs. (\ref{fig1},\ref{fig2})
(see also Fig. \ref{fig4}). Some technical details about how
the above integrals are handle can be found in Sec. IV of Ref.
\cite{se92}.

For a large antidot, the electronic density is constant everywhere
apart from a narrow region along the border of the hole.
Following the method outlined in Ref. \cite{li97},
it is easy to show that $E_3$ yields the classical
hydrodynamics dispersion relation for edge waves, namelly
$E_3 =   \omega(q) \sim \sqrt{2 \, n_s \,q\,\ln\,(q_0/q)}$.
It is also worth to notice that the induced (or transition) density
associated to the operator $Q_L$ has the form \cite{li97}
\begin{equation}
\rho_{tr}(\vec{r})
\propto L \frac{1}{r^{L+1}}\,\rho'(r)
\label{eq11}
\end{equation}
that manifests the edge character of the excitation.

Figure (\ref{fig3}) shows the frequency of the $L$ = 1 mode as a
function of $n_s$ compared with the experimental points of \cite{zh92}.
For completeness, we have also plotted
the results corresponding to $L$ = 2.
We have checked that similar results are obtained using as potential
barrier the parabola $V_p(r) = m^* \omega_0^2 (R - r)^2/2$ for
$r \leq R$ with $\omega_0 = 1 H^*$. One can see that the agreement
with experiment is good. Furthermore, our calculation yields a
frequency of $\sim$ 68 cm$^{-1}$ for the $R$ = 180 nm, $n_s = 9
\times 10^{11}$ cm$^{-2}$ antidot, in good agreement with the
experimental findings of Ref. \cite{bo96}.

We have also studied the size dependence of the mode energy.
Fig. (\ref{fig4}) represents the electronic densities for antidots
of $R$ = 10, 15 and 20 $a_0^*$, and a surface density
$n_s$ = 0.2 $(a_0^*)^{-2}$. The frequencies of the
$L$ = 1 and 2 modes are shown in Fig. (\ref{fig5}) as a
function of $1/\sqrt{R}$.
They exhibit a distinct $R$ dependence, indicating a clear departure
from parabolicity of the confining potential, i.e., a physical
situation where the generalized Kohn theorem does not apply.

If the electronic density $\rho(r)$
is approximated by a quasi-step function,
an analytical expression can be obtained for $E^2_3$.
Proceeding as in Ref. \cite{li97} one gets:
\begin{equation}
E^2_3= \pi n_s \frac{L(L+1)}{R^2}+ 4 n_s\frac{L}{R}\,\left[
\frac{1}{2}\,\, ln\,\left(\gamma \frac{R}{a}\right)
+1-\sum^L_{m=1}\frac{ 1}{2m-1} \right]  \,\,\, ,
\label{eq12}
\end{equation}
where $a$ represents the width of the quasi-step function, and the
precise value of $\gamma$ depends on the way the electronic density
goes to zero \cite{gi94}.
This equation tells one that the frequencies
have a $1/\sqrt{R}$ linear dependence if
the Coulomb energy contribution dominates.
For reasonable values of $\gamma/a$, it happens
for any realistic value of $R$.

Equation (\ref{eq12}) is too crude an approximation because of the
quasi-step function density used to get it. Much
better results can be obtained if
Eqs. (\ref{eq5}-\ref{eq9})
are evaluated in the Thomas-Fermi approximation,
i.e. $\tau(r) = \pi \rho^2(r)$, $\lambda(r) = \tau(r)/2$
 \cite{li97}, after
having fitted the KS density to a generalized Fermi function of the kind
\begin{equation}
\rho(r)= n_s \,\left (1\,-\,\frac{1}{1+e^{(r-R)/a}} \right)^{\nu}
\,\,\, .
\label{eq13}
\end{equation}
For $n_s$ = 0.2 $(a_0^*)^{-2}$, the KS densities are well reproduced
on average taking $\nu$ = 1.1 and $a$ = 0.38 $a_0^*$
(see Fig. \ref{fig4}). The mode
frequencies are represented by the lines in Fig. (\ref{fig5}).

In conclusion, we have shown that Density Functional Theory is able to
reproduce the zero magnetic field frequency of antidot edge modes
in quite a similar way as it does for quantum dots.
Although a satisfactory description of the collective spectrum of
antidots can be achieved using a magnetoplasmon approach \cite{wu93},
an alternative method based on a more microscopic approach such as
DFT is needed to discuss other interesting
problems, such as edge reconstruction and the formation of
compressible and incompressible strips at the antidot edge
\cite{bo96}.

It is a pleasure to thank Ricardo Mayol and Francesc Salvat
for useful discussions.
This work has been performed under grants PB95-1249 and PB95-0492
from CICYT, Spain, and GRQ94-1022 from Generalitat of Catalunya. A.E.
acknowledges support from the Direcci\'on General de Ense\~nanza
Superior (Spain).

\begin{figure}
\caption{ Electronic densities as a function of
$r$  for  antidots of $R$ = 7.5 $a_0^*$ and $n_s$ = 0.05, 0.1,
0.2, 0.3 and 0.4 (a$^*_0$)$^{-2}$.
Also shown are the jellium densities (dotted lines).}
\label{fig1}
\end{figure}
\begin{figure}
\caption{ KS potential $V(r)$  as a function of
$r$  for  antidots of $R$ = 7.5 $a_0^*$ and $n_s$ = 0.1
and 0.4 (a$^*_0$)$^{-2}$ (solid lines). The dashed lines
represent $V(r)$ without the $V_p(r)$ contribution.
The horizontal solid lines indicate the Fermi level, and the
vertical dotted line, the radius of the antidot.}
\label{fig6}
\end{figure}
\begin{figure}
\caption{Same as Fig. 1 for $R$ = 18.35 $a_0^*$ and $n_s$ =
0.86 ($a^*_0$)$^{-2}$. The dashed line density has been obtained
setting $\omega_0$ = 0.}
\label{fig2}
\end{figure}
\begin{figure}
\caption{Mode frequency as a function of the electron surface density
for $L$ = 1 and 2. The crosses are experimental data from Ref. [2],
and the lines are drawn to guide the eye.}
\label{fig3}
\end{figure}
\begin{figure}
\caption{ Electronic densities as a function of
$r$  for  antidots of $R$ = 10, 15 and 20
 $a_0^*$, and $n_s$ = 0.2 (a$^*_0$)$^{-2}$.
Also shown are the jellium densities (dotted lines).
The thin solid line represents the parametrized density,
Eq. [13], for $R$ = 20 $a_0^*$. }
\label{fig4}
\end{figure}
\begin{figure}
\caption{$L$ = 1 and 2 mode frequencies as a function of $R^{-1/2}$ for
$n_s$ = 0.2 (a$^*_0$)$^{-2}$. From right to left, the dots correspond
to $R$ = 7.5, 10, 12.5, 15, 17.5, and 20  $a_0^*$. The lines
represent the results obtained using the density Eq. [13].}
\label{fig5}
\end{figure}
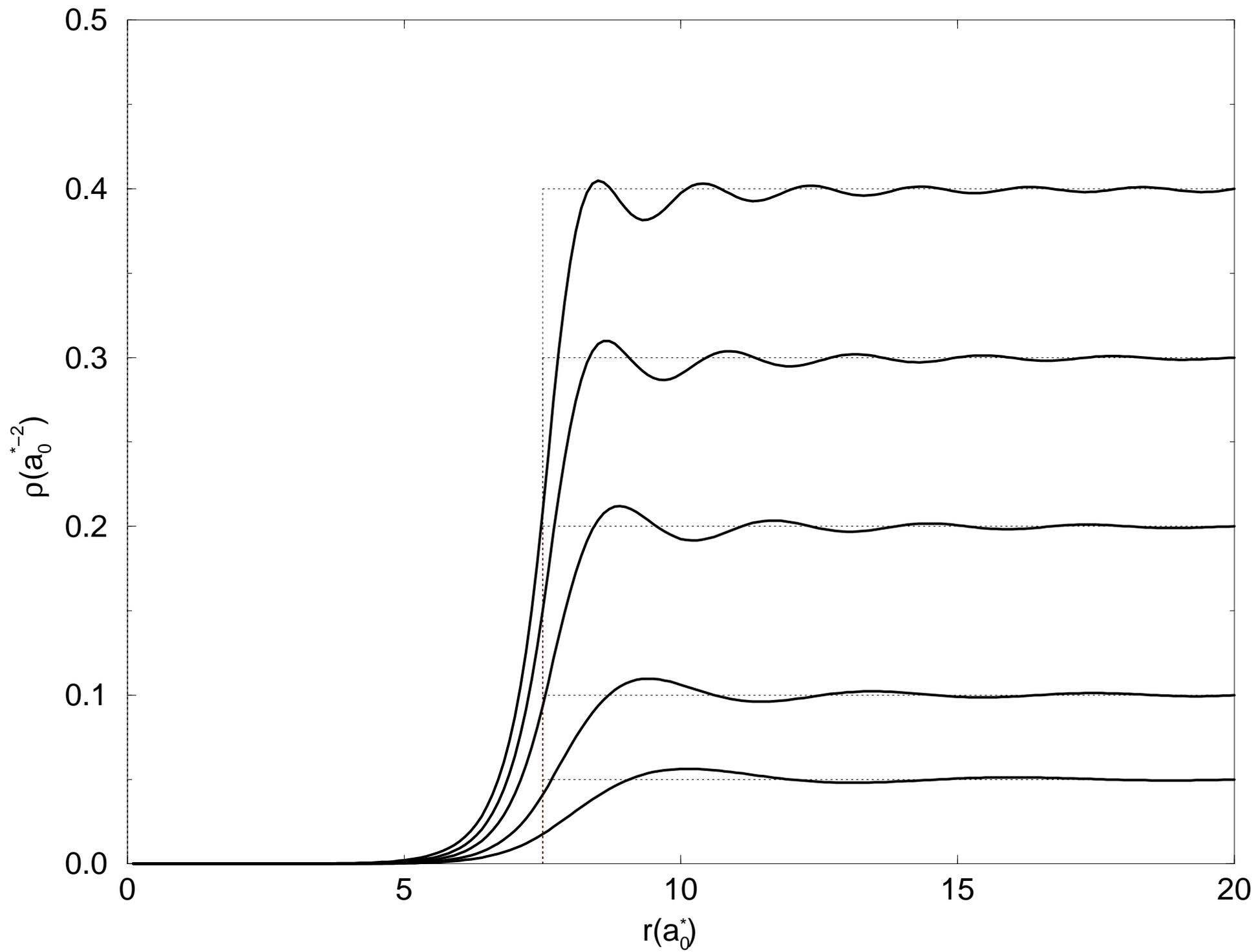
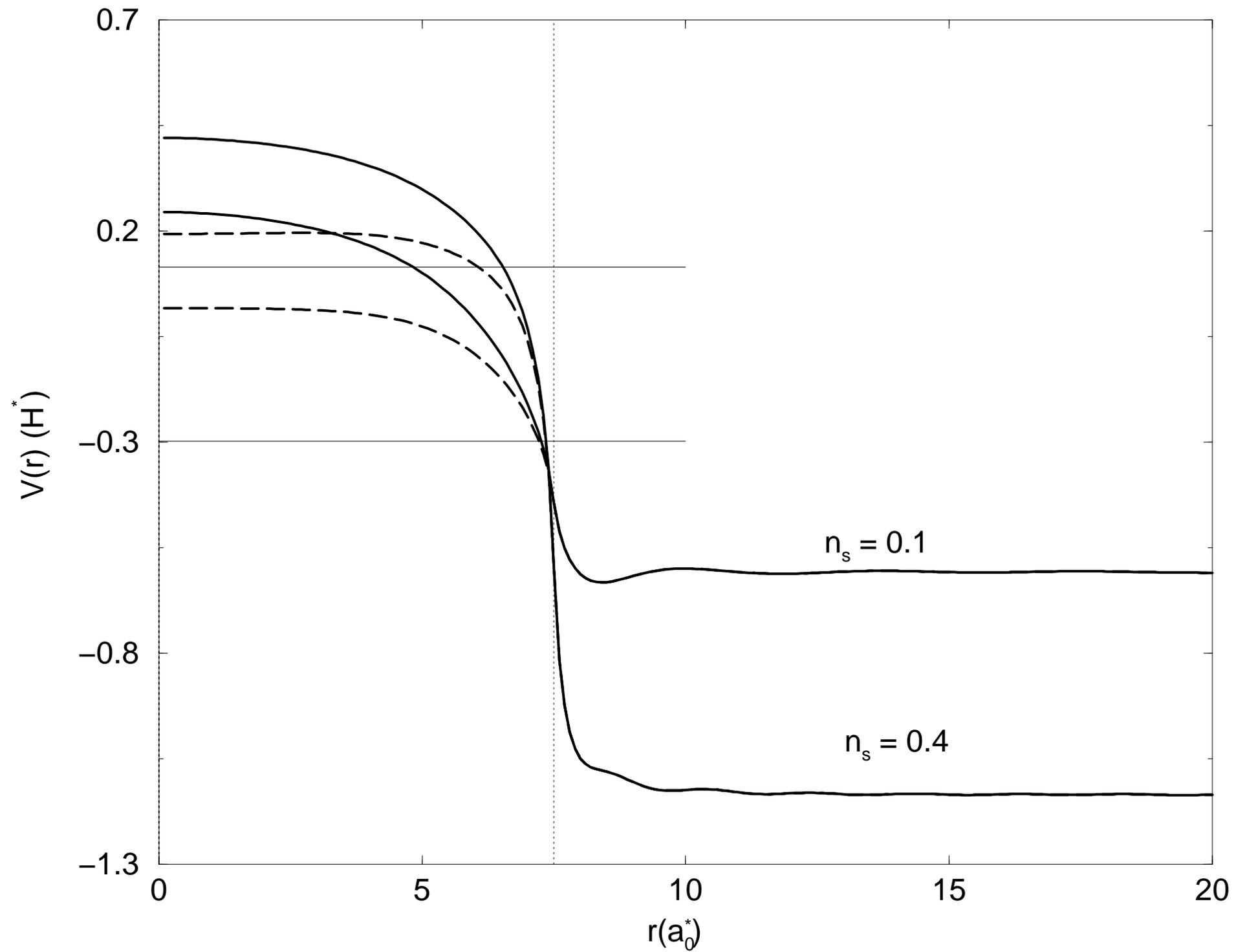
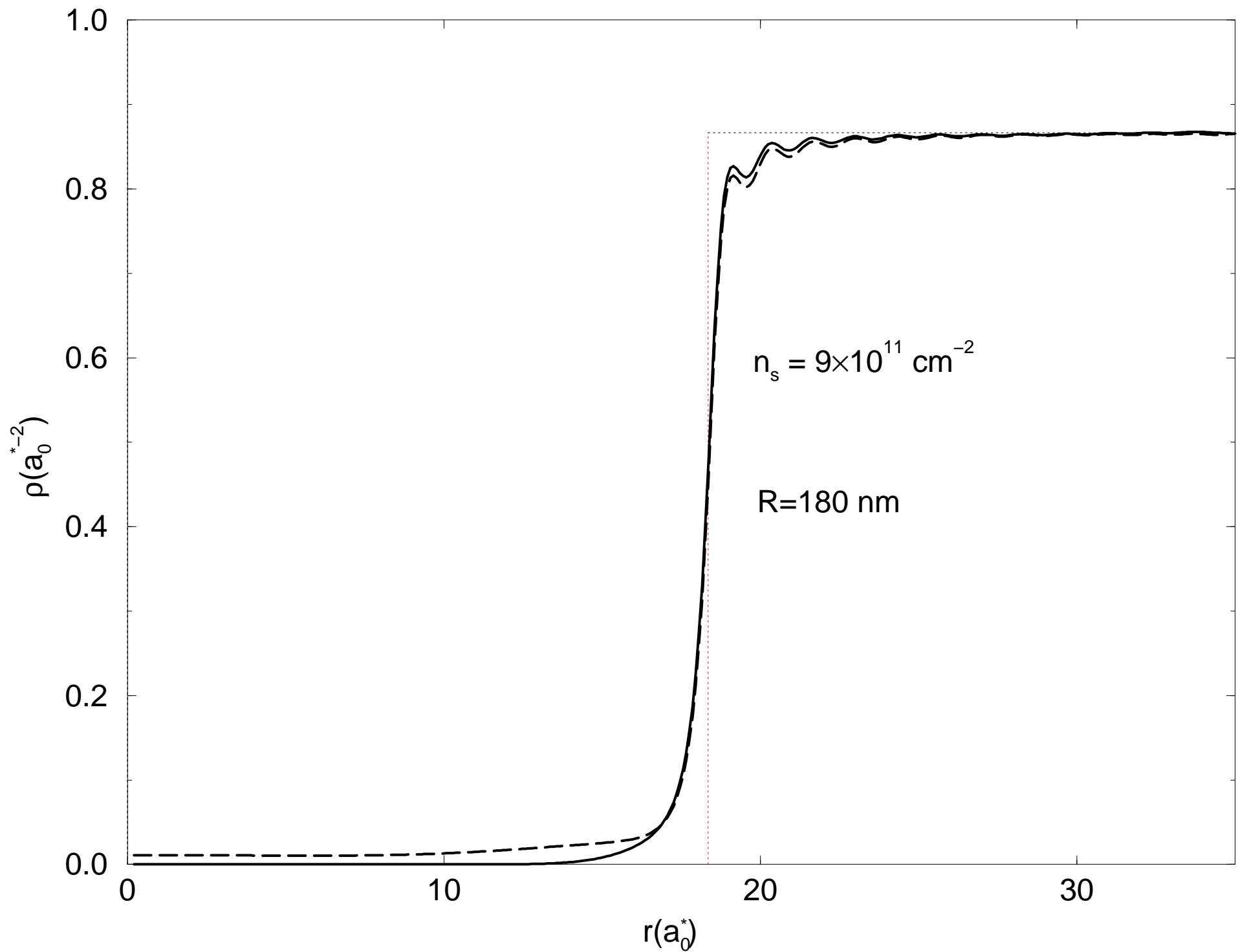
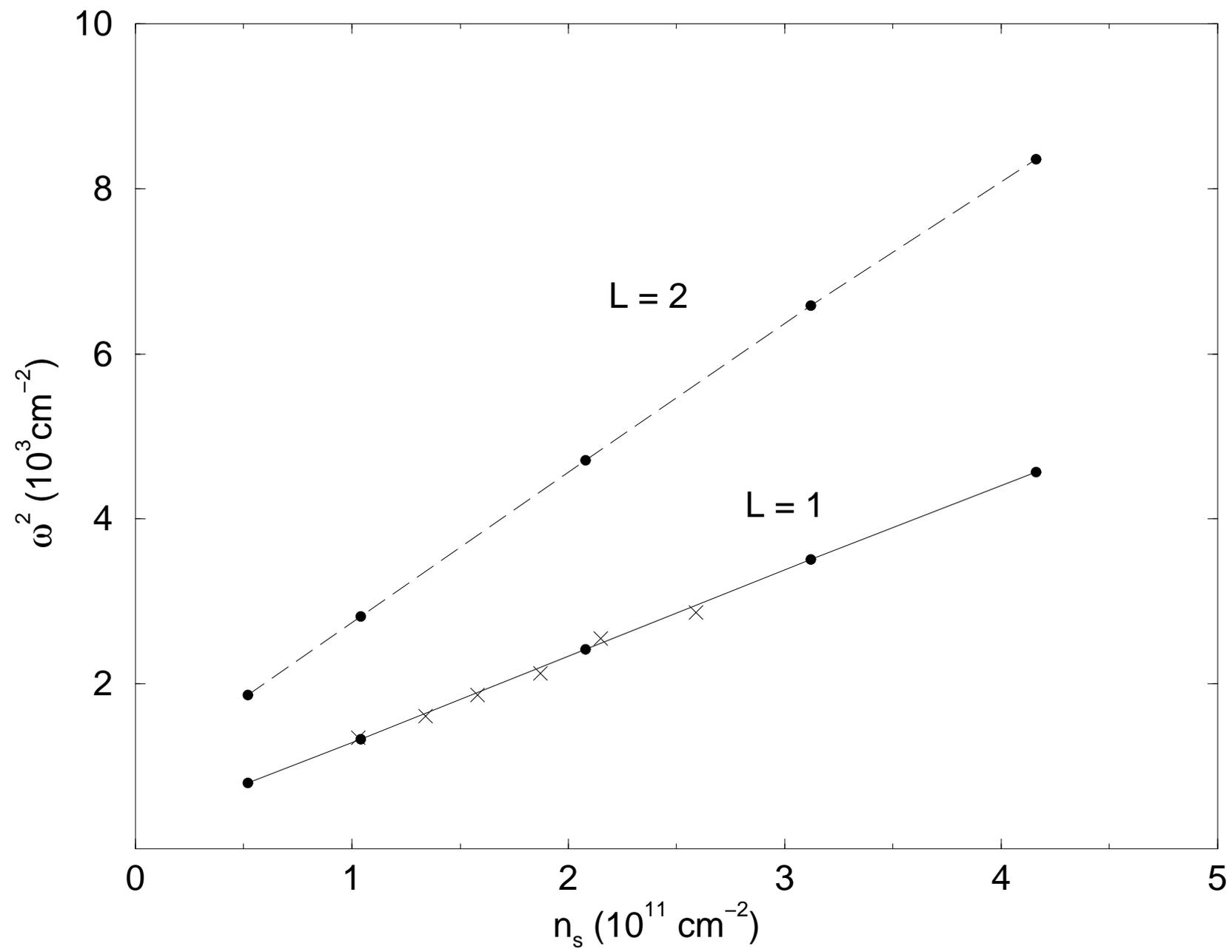
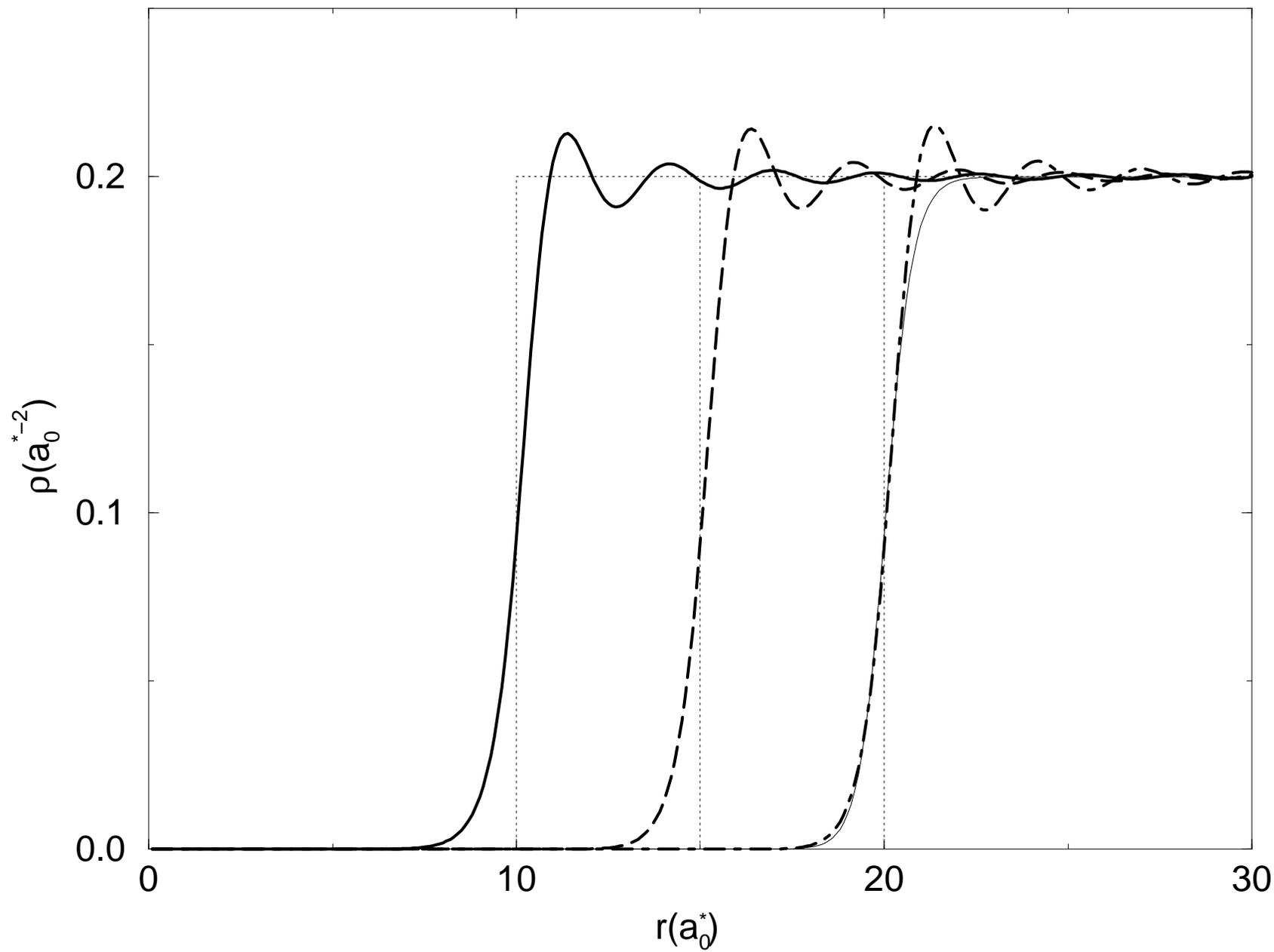
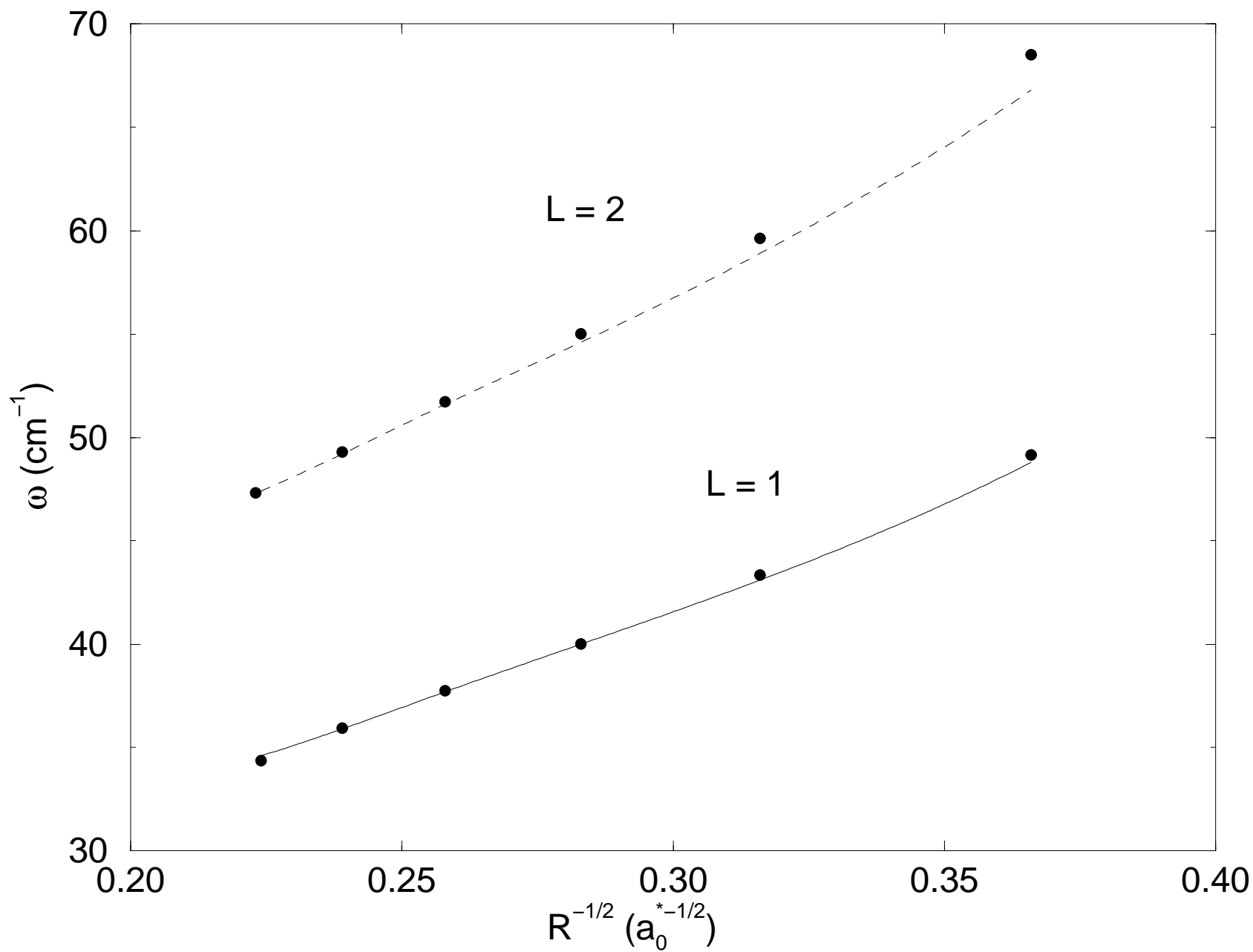
\end{document}